\documentclass[a4paper,11pt]{article}
\usepackage{pos}
\usepackage{braket}
\usepackage{graphics}
\usepackage{booktabs}
\usepackage{mathtools}
\usepackage{epstopdf}
\graphicspath{{./figs/}}
\usepackage{cleveref} 
\usepackage{wrapfig}
\newcommand{\mrgi}{M_{\text{RGI}}}
\newcommand{\msb}{\overline{\text{MS}}}
\newcommand{\mbar}{\overline{m}}

\newcommand{\eq}[1]{eq.~\eqref{#1}}
\title{Investigation of the Perturbative Expansion of Moments of Heavy Quark Correlators for $N_f=0$}
\ShortTitle{Heavy Quark Moments}

\author*[a,b]{Leonardo Chimirri}
\author[a,b]{Rainer Sommer}

\affiliation[a]{Humboldt Universit\"at zu Berlin, Institut f\"ur Physik \& IRIS Adlershof,\\ Zum Gro{\ss}en Windkanal 6, 12489 Berlin, Germany}

\affiliation[b]{DESY, Platanenallee 6, 15738 Zeuthen, Germany}

\emailAdd{leonardo.chimirri@desy.de}
\emailAdd{rainer.sommer@desy.de}
\abstract{The QCD-coupling is a necessary input in the computation of many observables, and the parametric error on input parameters can be a dominant source of uncertainty. The coupling can be extracted by comparing high order perturbative computations and lattice evaluated moments of mesonic two-point functions with heavy quarks, which provide a high energy scale for perturbation theory. The truncation of the perturbative series is an important systematic uncertainty.\\
We report on our attempt to study this issue by measuring pseudo-scalar two-point functions in volumes of $L=2\, \text{fm}$ with twisted-mass Wilson fermions in the quenched approximation. We use full twist, the non-perturbative clover term and lattice spacings down to $a=0.015\,\text{fm}$ to tame the sizable discretization effects. Our preliminary results indicate that either higher order perturbative corrections 
or the continuum limit are not  under sufficient control despite our small lattice spacings and quark masses
extending beyond $2\,m_{\text{charm}}$.}
\FullConference{%
 The 38th International Symposium on Lattice Field Theory, LATTICE2021
  26th-30th July, 2021
  Zoom/Gather@Massachusetts Institute of Technology
}

\begin{document}
\maketitle

\section{Introduction and Motivation}
The running QCD-coupling $\alpha(\mu)$ is one of the fundamental parameters of the Standard Model and its precise knowledge is needed for many  predictions in phenomenology. For instance, the parametric error on $\alpha(\mu)$ is crucial in partial widths of the Higgs boson such as $H\,\rightarrow\,b\bar{b}$, $H\,\rightarrow\,gg$~\cite{Almeida:2013jfa}. Lattice computations of the coupling are today among the most competitive ones \cite{Aoki:2021kgd,Zyla:2020zbs,DelDebbio:2021ryq}. It is mandatory to check that all sources of errors in their determinations are under control. \\
We focus here on the so called ``moments method'', introduced in~Refs.~\cite{Bochkarev:1995ai,HPQCD:2008kxl}, in which concurrent high order perturbative and lattice non-perturbative knowledge of these observables is used to extract the coupling. Here we want to study the main systematic errors present in this approach, namely the truncation of the perturbative series, as well as the reliability of the continuum extrapolation for these observables. \\
On the lattice side, one wants the typical scale of the observable, in this study given by the heavy quark mass $m_h$, as well as all other relevant scales to be far from the cutoff, while the lattice size $L$ has to be large enough to avoid finite size effects. At the same time we need $m_h\gg\Lambda$ for the coupling to be small, so that perturbation theory provides a usable asymptotic expansion. These conditions are best summarized as
\begin{equation}
	L^{-1}\ll \Lambda \ll m_h\ll a^{-1}\,,
\end{equation} 
where a limit on $L/a$ is set by computational resources. Since it is notoriously difficult to control cutoff effects in this context, our study is done fully in the quenched model, where it is much easier to simulate at small lattice spacings. 

\section{Definition of the Moments}
The definition of the moments in the continuum and in Euclidean spacetime is given by 
\begin{equation} \label{eq:mom_cont_def}
	\mathcal{M}_n(\mrgi) = \int_{-\infty}^{+\infty}\text{d}t\,t^n \int \text{d}^3x\, m_h^2\left\langle J^{\dagger}(x)J(0)\right\rangle\,,\quad J(x)=i\overline{\psi}_h(x)\gamma_5 \psi_{h'}(x)\,,
\end{equation}
where $m_h=m_{h'}$ denotes the mass of a heavy-quark doublet\footnote
	{We have  two different, mass-degenerate flavors in order not to have disconnected diagram contributions. 
	} 
which is one-to-one with $\mrgi$, the renormalization group independent quark mass. Note that using the pseudo-scalar density is a choice, and phenomenological studies use also other $\gamma$-structures, which are, however, affected by larger statistical noise. The mass factor is introduced to make $\bar{J}(x)=m_h\,J(x)$ renormalization group invariant if $Z_PZ_m=1$ holds, such as for the twisted mass fermions (at full twist) employed in this study. The moments have a weak coupling expansion (in terms of $\msb$-renormalized parameters)
\begin{equation} \label{eq:mom_pert_exp}
	\mathcal{M}_n^{(\text{PT})}(\mrgi)=
	\mbar_{\msb}(\mu)^{4-n}
	\sum_{i=0}^3\,c_n^{(i)}(\mu/\mbar_{\msb}(\mu))\,\alpha^i_{\msb}(\mu)+ \mathcal{O}(\alpha^{4})\,,
\end{equation}
where the coefficients $c_n^{(i)}(\mu/\mbar_{\msb}(\mu))$ are know for $i=0,\,1,\,2,\,3$ \cite{Chetyrkin:1997mb,Maier:2009fz} from perturbation theory (PT). Note, in $\mathcal{M}_n^{(\text{PT})}$ there is a spurious $\mu$-dependence originating from the truncation of the perturbative, asymptotic series. The variation of this very scale around the physical scale is often used to assess the size of the truncated terms~\cite{Dehnadi:2015fra,DelDebbio:2021ryq}.\\
In terms of bare quantities on a finite lattice with periodic boundary conditions in space and open boundary conditions (to avoid topological freezing, see \cite{Luscher:2011kk}) in time, \eq{eq:mom_cont_def} becomes
\begin{align} \label{eq:lat_moments}
	&\mathcal{M}^{(\text{lat})}_n(\mrgi,\,a)=
	\lim_{T,L\to\infty}2a\sum_{t=x_0^{\text{src}}}^{\delta}t^n\left(\frac{a}{L}\right)^3\,a^3\sum_{{\bf x,\,y}=0}^{L-a} \mu_{tm}^2 \left\langle J^{\dagger}(t, {\bf x}) J(0, {\bf y})\right\rangle\,,\quad\text{with}
	\\ 
	&\lim_{a\to0}\mathcal{M}^{(\text{lat})}_n(\mrgi,\,a)=
	\mathcal{M}_n(\mrgi)\,, \label{eq:cont_lim_mom_def}
\end{align}
where $\mu_{tm}$ is the bare, twisted mass, $\delta$ is a cutoff introduced to avoid values of $t$ affected by the boundary at $x_0=T$ and the factor $2$ accounts for negative times present in the continuum definition (the PS-PS correlator is time-symmetric). Independence of results on $\delta$ is checked.

\section{Methodology}
\begin{figure}
	\centering
	\includegraphics[width=1.0\textwidth ]{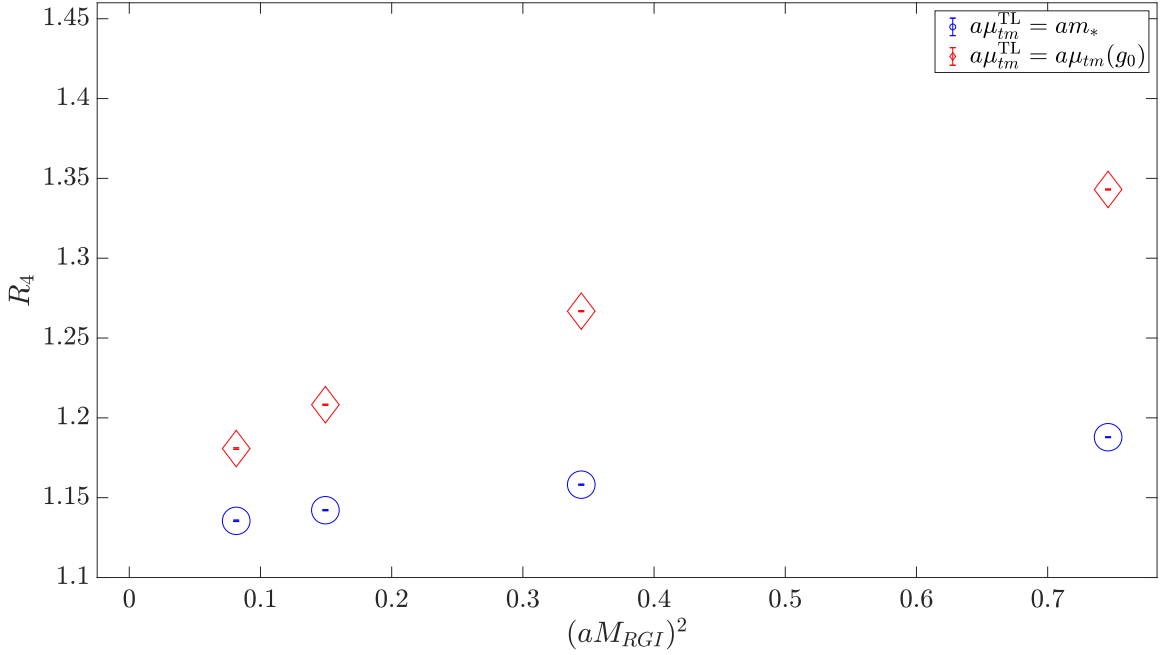} 
	\caption{Improvement seen in the continuum approach of $R_4$ for $z=9$ when setting the scale of the analytically computed, finite lattice spacing tree-level to the relevant mass scale.}
	\label{fig:TL_scale_improvement_2p3216}
\end{figure}

$\mathcal{M}_n^{(\text{lat})}$ of \eq{eq:lat_moments} is finite in the continuum limit and needs no further renormalization, as long as the short distance divergence is integrable, which is the case\footnote
{
	From an OPE analysis one finds that
	\begin{equation}
		G(t)\coloneqq \int \text{d}^3x\, m_h^2\left\langle J(x)J(0)\right\rangle
		\overset{t\to0}{\sim} \frac{1}{|t|^3}cnst.\left(\overline{g}^2(1/|t|)\right)^{\gamma_0/\beta_0}\left(1+\mathcal{O}(\overline{g}^2(1/|t|))\right)\,,
	\end{equation}
	denoting by $\gamma_0$ the leading order pseudoscalar anomalous dimension and by $\beta_0$ the leading $\beta$-function coefficient.
} 
for $n\ge4$. We define a line of constant physics by fixing the value of the RGI-mass in units of the gradient flow scale $\sqrt{8t_0}=0.463(3)\,\text{fm}$ \cite{Luscher:2010iy}, i.e. we tune the twisted mass parameter so that
\begin{equation}
z=\sqrt{8t_0}\mrgi=
\frac{\sqrt{8t_0}}{a}\frac{\mrgi}{\mbar_{\text{SF}}(\mu)}a\mu_{tm}\left(Z^{\text{SF}}_P(a\mu,g_0)\right)^{-1}\,,
\end{equation}
is equal to some chosen value and constant as we take the continuum limit. In the above, $\mrgi/\mbar_{\text{SF}}(\mu)$ and $Z_P^{\text{SF}}(a\mu,g_0)$ are taken from a reanalysis of the data of Ref.~\cite{Capitani:1998mq} for 
\begin{wraptable}{r}{3.5cm}
	\caption{Mass values in units of $\sqrt{8t_0}$ and of the quenched RGI-charm mass.}
	\label{table:masses}
	\begin{tabular}{c c}  
		\hline\toprule
		z&$\mrgi/\mrgi^{\text{charm}}$\\
		\midrule
		13.5&3.48\\
		\midrule
		9&2.32\\
		\midrule
		6&1.55\\
		\midrule
		4.5&1.16\\
		\midrule
		3&0.77\\
		\bottomrule
	\end{tabular}
\end{wraptable}
$\mu=L_{\text{ref}}^{-1}$, where $L_{\text{ref}}$ is defined as $\overline{g}^2(L_{\text{ref}})=2.4484$; finally $\sqrt{8t_0}/a$ can be measured with accuracy below the level of other errors. Equating \eq{eq:cont_lim_mom_def} with \eq{eq:mom_pert_exp}, one can extract the $\msb$-parameters at some renormalization scale. Since we are mostly interested in the coupling, we will work with dimensionless observables in which non-logarithmic, i.e. rather strong, dependence on $\mbar_{\msb}\,$ drops out.\\
Measurements are carried out for the range of masses found in table \ref{table:masses} where the quenched charm mass is taken to be $\mrgi^{\text{charm}}=1.684(60)\,\text{GeV}$ \cite{Rolf:2002gu}. In this way, the scale dependence of the coupling can be studied. Let us mention that the larger $n$ is, the more the integral/sum is dominated by large, non-perturbative, distances so that the values most used in literature are $n=4,\,6,\,8,\,10$. \\
Two further modifications of the moments are introduced. First, the tree-level moments,
\begin{equation}
	\mathcal{M}_n^{\text{TL}}(a\mu_{tm},L/a)\coloneqq \mathcal{M}^{(\text{lat})}_n\big|_{g_0=0}\,,
\end{equation}
 are analytically computed, so that the ratios 
\begin{equation}
	R_n(a\mrgi,\,z)= 
	\begin{cases}
		&\frac{\mathcal{M}_n(a\mrgi,z)}{\mathcal{M}_n^{\text{TL}}(a\mu_{tm}^\text{TL},L/a)}\,,\quad n=4\\
		&\left(\frac{\mathcal{M}_n(a\mrgi,z)} {\mathcal{M}_n^{\text{TL}}(a\mu_{tm}^\text{TL},L/a)}\right)^{\frac{1}{n-4}}\,,\quad n=6,\,8,\,10,\,
	\end{cases}		
\end{equation}
are formed. In this way the leading lattice artefacts of $\mathcal{O}(a^2)$  are suppressed by a factor of $\alpha$ (up to logs, see \cite{Husung:2019ytz}).
Formally, all we need is $a\mu_{tm}^\text{TL} = a m +\mathrm{O}(g^2)$ with $m$ any definition of the mass. At first sight one tends to set $a\mu_{tm}^\text{TL}=a\mu_{tm}(g_0)$ where $a\mu_{tm}(g_0)$ is the bare twisted mass of the simulation at coupling $g_0$. This choice is represented by the red diamonds in fig.~\ref{fig:TL_scale_improvement_2p3216}. A much better choice is to set  $a\mu_{tm}^\text{TL}=am_{\ast}$, where $m_{\ast}$ is a renormalized mass and we choose $m_{\ast}=\mbar_{\msb}(m_{\ast})$. We use it in the following and note that it  helps a lot in taking the continuum limit.\\
For dimensionful moments with $n>4$ the above ratio is elevated to some $n$-dependent power so that they all have mass dimension $d=-1$. In the case of dimensionful $R_n$, one can then take ratios of normalized moments and thus study the set of dimensionless observables
\begin{equation} \label{eq:dimless_moms}
	\mathcal{R}_n(a\mrgi,z) =
	\begin{cases}
		&R_4(a\mrgi,z)\,,\quad n=4,\\
		&\frac{R_n(a\mrgi,z)}{R_{n+2}(a\mrgi,z)}\,,\quad n=6,\,8\,.\\
	\end{cases}
\end{equation}
\section{Lattice Setup}
\begin{figure}
	\centering
	\includegraphics[trim=60 0 60 40, clip, width=0.94\textwidth ]{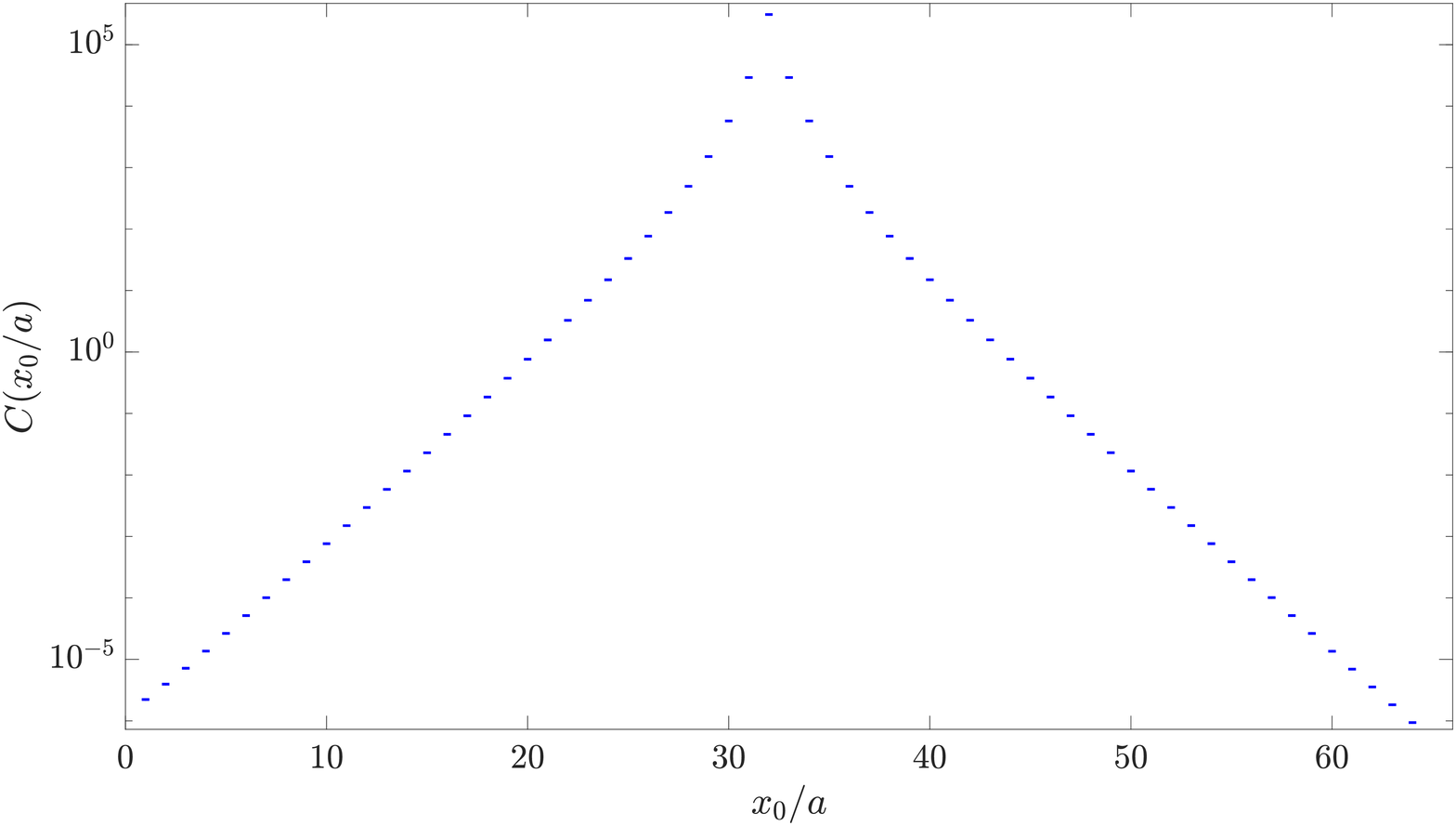} 
	\caption{$C(x_0)=\sum_{{\bf x,\,y}=0}^{L-a}\left\langle J^{\dagger}(x_0, {\bf x}) J(0, {\bf y})\right\rangle$. No asymmetry can be resolved within errors around the source time slice; $z=6$, $\beta= 6.7859$.}
	\label{fig:asym_source_M155}
\end{figure}

Let us give a brief overview of the lattice setup details. We use a plaquette action to generate pure gauge ensembles and a Wilson, mass-degenerate fermion doublet with a twisted mass term. We set $c_{\text{SW}}$ to its non-perturbative value, taken from \cite{Luscher:1996ug} in order to reduce ambiguities in the point of full twist and to avoid second order corrections of the Pauli-term in the Symanzik effective theory. Both might lead to enhanced cutoff effects. We work at full twist by tuning $\kappa$ to its critical value \cite{Luscher:1996ug} independent of the quark mass.  With units \cite{Sommer:1993ce} of $r_0=0.49\,\text{fm}$, the typical size of our lattice is $L\simeq2\,\text{fm}$ for the spatial extent and $T\simeq6\,\text{fm}$ for the time extent, where open boundary conditions are imposed. Sources are placed $\sim1\,\text{fm}$ away from the time boundary and the absence of boundary effects was checked by monitoring the correlator around the source time-slice (see \cref{fig:asym_source_M155}). The correlator is symmetric around the source time-slice with no significant deviation  seen up to very close to the boundary. In the example in \cref{fig:asym_source_M155}, for instance, an asymmetry can be resolved only at 4-5 lattice spacings distance from the boundary.
Random U(1)-noise sources are used for the estimate of the two-point function, with typically $N_S\simeq16$ noise vectors. The autocorrelation analysis is carried out with the $\Gamma$-method \cite{Wolff:2003sm}.\\
\begin{table}[!htb] 
	\caption{Gauge configuration details: $l=L/a$, $t=T/a$, sft-ensembles are from \cite{Husung:2017qjz}.}
	\centering
	\label{table:gauge_runs}
	\resizebox{0.75\columnwidth}{!}
	{
		\begin{tabular}{c c c c c c c}
			\hline \toprule
			Run Name& $\beta$ &  $l\times t$& $N_{\text{cnfg}}$ & $t_0/a^2$ &$a[{\rm fm}]$ &$\tau_{\text{int}}(t_0)[\text{cfg}]$\\
			\midrule[1.5pt]
			q\_beta616&6.1628 & $32\times96$ & 128&5.376(10)& 0.071& 0.78\\
			\midrule
			q\_beta628&6.2885& $36\times108$ &  137 &7.790(22)& 0.059 & 1.37\\
			\midrule
			q\_beta649&6.4956 &$48\times144$&   109  &13.778(51)& 0.044 & 1.55\\
			\midrule[1.5pt]
			sft4 &6.7859 & $64\times192$&  200 &29.39(10)& 0.030 & 1.00 \\			
			\midrule
			sft5&7.1146&$96\times320$& 80 &67.74(23)& 0.020 &0.55\\
			\midrule
			sft6 & 7.3600 &$128\times320$&  98 &124.21(91) & 0.015 & 1.03\\			
			\midrule
			sft7 & 7.7 & $192\times480$&  55 & &$$0.010&\\			
			\bottomrule			
		\end{tabular}
	}
	\label{tab:gauge_runs}
\end{table}

\section{Results}
In this Proceedings we report our preliminary results on $R_4$ and on the $\Lambda$-parameter extracted from this moment. Results from higher moments  will be discussed in a future publication.
\subsection{Continuum Extrapolations}
\begin{figure}
	\centering
	\includegraphics[width=1.0\textwidth ]{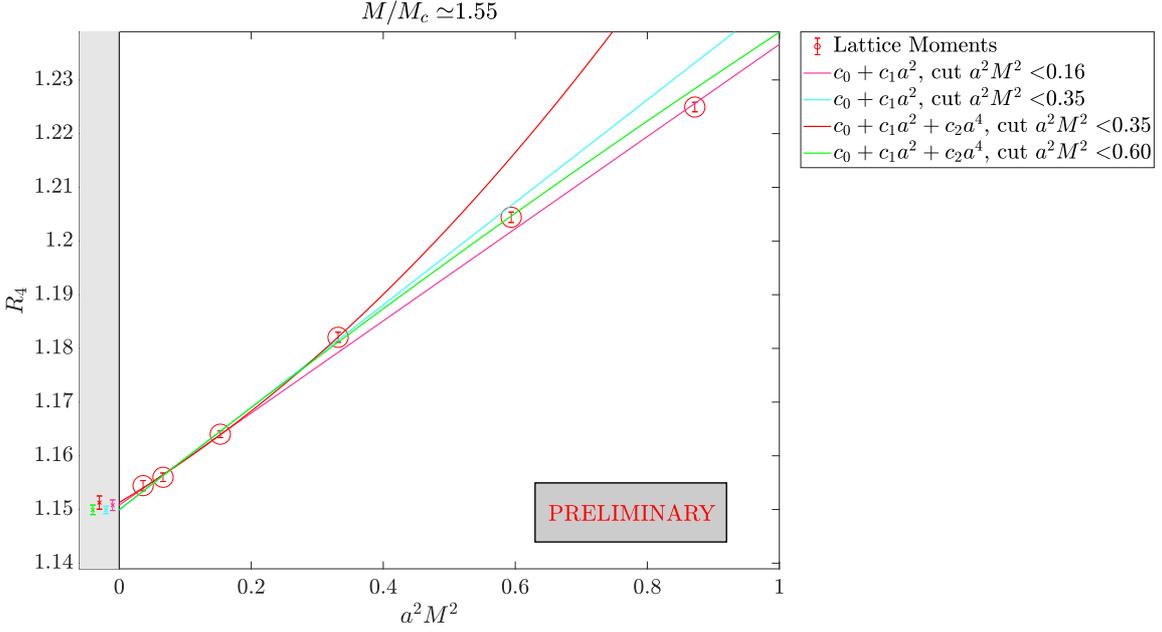} 
	\caption{Continuum limit for $z=6$, or $M\simeq1.55M_c$.}
	\label{fig:r4_m1p5478}
\end{figure}

\begin{figure}
	\centering
	\includegraphics[trim=90 0 60 30, clip, width=1.0\textwidth ]{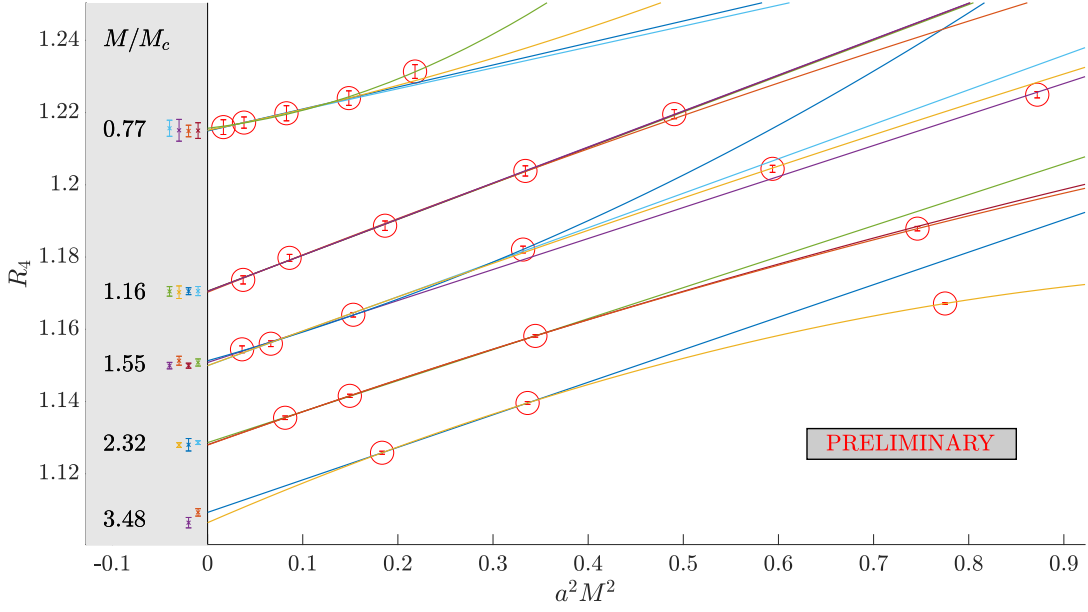} 
	\caption{Camparison of continuum limit for different masses.}
	\label{fig:multiple_masses}
\end{figure}

The continuum extrapolations vary in quality depending on the observable and mass. In \cref{fig:r4_m1p5478} one such continuum limit is shown for several fit ans\"atze. Here and in the following, only fits with $\chi^2/\text{dof}<2$ are considered. The extrapolated value for each fit is shown in the gray band on the left hand side. The discretization errors scale roughly like $a^2M^2$ as seen in fig.~\ref{fig:multiple_masses}. Higher quark masses require a better resolution and are  computationally more challenging. Continuum limit fits are shown for all ans\"atze which satisfy our cut on $\chi^2$. We are not able to take a significant continuum limit for the highest mass $\mrgi/\mrgi^{\text{charm}}\simeq3.48$, but hope we will be able to do so once data on the smallest lattice spacing (ensemble sft7) is available.
At the larger masses, the extrapolation from the last computed point to the continuum limit is still rather significant. A correct extrapolation therefore depends
on having (approximately) the correct extrapolation function. As a precaution we therefore add to our continuum extrapolation result a further error, given by half of the distance between the extrapolated result and the point closest to the continuum. We will discuss the continuum limit further in sect.~\ref{s:cont}.\\

\subsection{Coupling Values}
The observables defined in \eq{eq:dimless_moms} all have a perturbative expansion
\begin{equation} \label{eq:Rn_pt_expansion}
	\lim_{a\to0}\mathcal{R}_n(a\mrgi,z) =
	\sum_{i=0}^{3}d_i(s)\,\alpha_{\msb}^i(\mu_s)+\mathcal{O}(\alpha^4)\,,\quad
	\mu_s=s\mbar_{\msb}(\mu_s)\,,
\end{equation}
where the coefficients $d_i(s)$ can be given in terms of the $c_n^{(i)}(\mu/\mbar_{\msb}(\mu))$ in \eq{eq:mom_pert_exp} and $s$ is a scale parameter chosen of $\mathcal{O}(1)$. For a given $s$, there is a unique  value $\mu_s$. The parameter $s$ can be varied to probe the behavior of PT, namely the size of the truncated terms. By inverting \eq{eq:Rn_pt_expansion} we obtain $\alpha_{\msb}(\mu_s)$, which can be run from $\mu_s$ to infinite energy to obtain the $\Lambda$-parameter.

\subsection{$\Lambda$ Parameter} \label{sec:lambda_parameter}
\begin{figure}
	\centering
	\includegraphics[trim=100 0 120 40, clip, width=1.0\textwidth ]{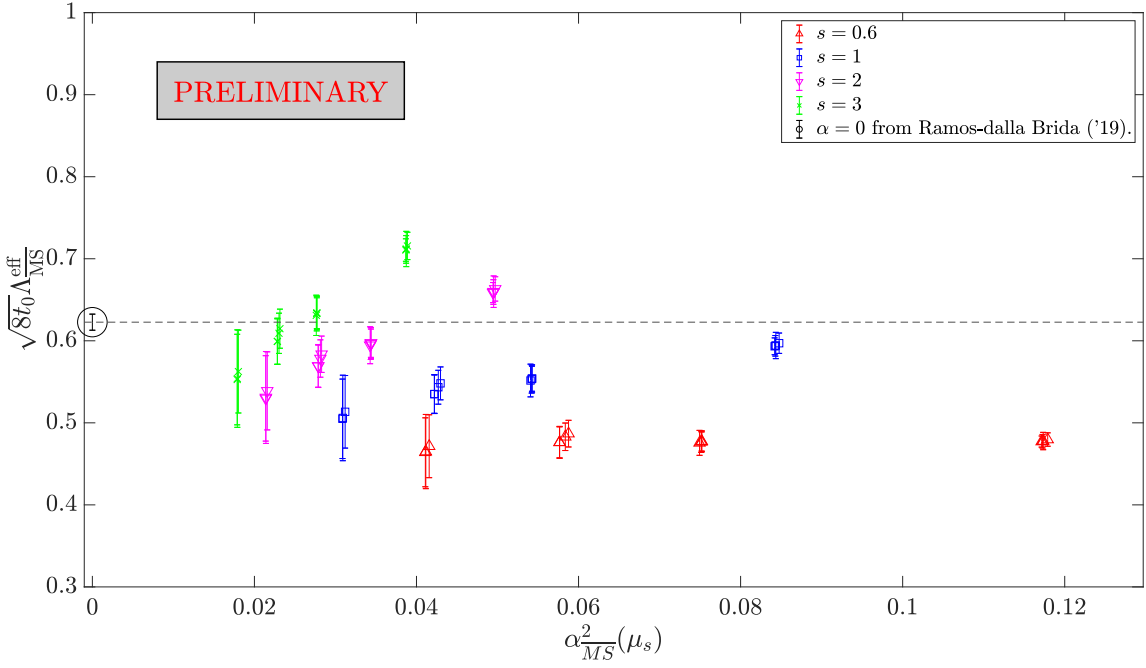} 
	\caption{Results for $\Lambda^\mathrm{eff}_{\msb}$ computed from $R_4$. Each value of the scale parameter $s$ is in a different color and shape, while nearby points are the result from different fit Ans\"atze. The dashed line is plotted to guide the eye, and is an extension of the result for $\Lambda_{\msb}$ by \cite{DallaBrida:2019wur}, which is in the limit $\alpha\to0$.}
	\label{fig:lambda_r4_all_s_extraerror}
\end{figure}

For any given $s$ and $z$, one can extract $\sqrt{8t_0}\Lambda_{\msb}$ from the ratio of the definition of the RGI-parameters using the perturbative beta function, $\beta_{\msb}(g)$, and quark mass anomalous dimension, $\tau_{\msb}(g)$,  \cite{Chetyrkin:1997dh,Vermaseren:1997fq,Herzog:2017ohr,Luthe:2017ttg} and by plugging in the value of $g_{\msb}$ obtained above:
\begin{align} 
	\frac{\sqrt{8t_0}\Lambda_{\msb}}{\sqrt{8t_0}\mrgi}= 
	&s\frac{(b_0 g_{\msb}(\mu_s)^2)^{-b_1/(2b_0^2)}} {(2b_0g_{\msb}(\mu_s)^2)^{-d_0/(2b_0)}}
	\exp\left\{-\frac{1}{2b_0g_{\msb}(\mu_s)^2}\right\} \times\\
	&\exp\left\{ -\int_{0}^{g_{\msb}(\mu_s)} \text{d} x \left[\frac{1-\tau_{\msb}(x)}{\beta_{\msb}(x)}+ \frac{1}{b_0x^3} -\frac{b_1}{b_0^2x}+\frac{d_0}{b_0x} \right] \right\}\,.
\end{align}
Resulting numbers are an estimator, $\Lambda^\mathrm{eff}_{\msb}$, of $\Lambda_{\msb}$. The estimator depends on the coupling from which it was extracted because we use a perturbative approximation both in \eqref{eq:Rn_pt_expansion} and in the RG functions. The latter is subdominant but the $\mathcal{O}(\alpha^4)$-uncertainty in $\mathcal{R}_n$ implies
\begin{equation}
	\Lambda_{\msb}^{\text{eff}}=
	\Lambda_{\msb}+\mathcal{O}\left(\alpha^2_{\msb}\,(\mu_s)\right)\,.
	\label{e:Lameff}
\end{equation}
In \cref{fig:lambda_r4_all_s_extraerror}, $\Lambda_{\msb}^\mathrm{eff}$
 is plotted against $\alpha^2(\mu_s)$  for different choices of $s$. Same-colored points close to one another result from different fit ans\"atze, while different colors indicate different values for the perturbative scale factor $s$.

\section{Discussion} \label{s:cont}
 
The results for $\Lambda^\mathrm{eff}_{\msb}$ depend strongly on $\alpha_{\msb}$ and on the scale parameter $s$. Taking the full variation with both of these parameters as an error estimate one
ends up with a very large error. In the past, in QCD with three or more light quark flavours \cite{Chakraborty:2014aca,Petreczky:2020tky,Petreczky:2019ozv,Nakayama:2016atf}, subsets of such variations have been taken to estimate the uncertainty. 

However, a systematic determination of $\Lambda_{\msb}$ from $\Lambda_{\msb}^\mathrm{eff}$ is
to fix $s$, reach the perturbative regime where the $\alpha^2$ term dominates in \eq{e:Lameff} and then extrapolate also in $\alpha$. One can see by eye that each set with fixed $s$ is compatible 
with a pure $\alpha^2$ correction and the different sets tend to extrapolate to an approximately consistent value which is, however, quite a bit below $\Lambda_{\msb}$ of
Dalla Brida and Ramos \cite{DallaBrida:2019wur}. The reason is not known at present. 

A first possibility is that none of our data are in the region where perturbation theory applies well and the observed agreement  with an $\alpha^2$ scaling is  accidental. There are other observables which require small $\alpha$ to approach the perturbative region \cite{DallaBrida:2018rfy}.

A second possibility is that our continuum extrapolation formulae are not quite adequate.  
 Indeed, our extrapolation in powers of $a^2$ is motivated by Symanzik effective theory (neglecting 
 the log-corrections originating from the 
 renormalization of the operators in the effective theory \cite{Husung:2019ytz}). However,
 Symanzik effective field theory is built 
 for correlation functions at long distances, not for the short distance region contained in the moments. It is plausible that for $n=6$ the short distance region is sufficiently suppressed, but for the here discussed -- and most perturbative -- moment $\mathcal{M}_4$, short distance effects result in $\log(a)$-enhanced cutoff effects \cite{Ce:2021xgd} -- at least at tree level. While our tree-level improvement removes part of those, as a next step we need to investigate whether  our continuum limit is really safe. To this end we want to perform a partial subtraction of the short distance region similar to the subtraction proposed in \cite{Ce:2021xgd} and also work towards developing a more systematic approach to the issue.

\paragraph{Acknowledgements}
This project has received funding from the European Union’s Horizon 2020 research and
innovation programme under the Marie Skłodowska-Curie grant agreement No. 813942.
\bibliographystyle{./JHEP}
\bibliography{./chimirri_HQmoments}
\end{document}